\documentclass[aps,twocolumn,groupedaddress]{revtex4-1}

\usepackage{graphicx}
\usepackage{siunitx}
\usepackage{sidecap}
\begin{document}

\title{Uniform coating of self-assembled non-iridescent colloidal nanostructures using Marangoni effects and polymers}

\author{Seung Yeol Lee$^{1,2}$}
\author{Hyoungsoo Kim$^{1,3}$}
\email[E-mail:]{hshk@kaist.ac.kr} 
\author{Shin-Hyun Kim$^{2}$}
\author{Howard A. Stone$^{1}$}
\email[E-mail:]{hastone@princeton.edu} 

\affiliation{$^1$Department of Mechanical and Aerospace Engineering, Princeton University, Princeton, NJ 08544, USA.}
\affiliation{$^2$Department of Chemical and Biomolecular Engineering, Korea Advanced Institute of Science and Technology (KAIST), Daejeon 34141, Republic of Korea.}
\affiliation{$^3$Department of Mechanical Engineering, Korea Advanced Institute of Science and Technology (KAIST), Daejeon 34141, Republic of Korea.}

\date{\today}

\begin{abstract}
Colloidal crystals exhibit structural color without any color pigment due to the crystals' periodic nanostructure, which can interfere with visible light. This crystal structure is iridescent as the resulting color changes with the viewing or illumination angle, which limits its use for printing or displays. To eliminate the iridescent property, it is important to make the packing of the colloidal nanoparticles disordered. Here, we introduce a drop-casting method where a droplet of a water-ethanol mixture containing monodisperse polymer-coated silica nanoparticles creates a relatively uniform and non-iridescent deposit after the droplet evaporates completely on a heated substrate. The uniformity is caused by a thermal Marangoni flow and fast evaporation effects due to the heated substrate, whereas non-iridescence is the outcome of short-range-ordered packing of nanoparticles by depletion attraction and friction effects produced by polymer brushes. We show that the colors of the final deposits from individual droplets remain unchanged while the viewing angle is varied under ambient light. We expect that the coating method is compatible with ink-jet printing and the uniformly coated self-assembled non-iridescent nanostructures have potential for color displays using reflection mode and other optical devices.
\end{abstract}


\maketitle

\section {INTRODUCTION}
It is well-known that the dried deposit from a coffee drop leaves a stain with a ring shape \cite{deegan1997capillary}. For example, when a droplet of a dilute suspension (3 wt.\%) of monodisperse silica nanoparticles (diameter $d$ = 228 nm) dries on a cover glass, the final deposit shows a green “coffee-ring stain”, as shown in Fig. 1(a). This green structural color is determined by the size and the packing or organization of the particles. During the drying process, slowly deposited silica nanoparticles form a close-packed, face-centered-cubic (fcc) structure in the vicinity of the contact line, which is induced by an evaporatively-driven outward flow~\cite{marin2011order}. The surface of the final deposit consists of an ordered hexagonal array of silica nanoparticles [Fig. 1(b)], which is confirmed by a two-dimensional fast Fourier transform (2d FFT) [inset of Fig. 1(b)], and the regular structure of the final deposit is confirmed by a cross-sectional image [Fig. 1(c)]. We note that the fcc structure of the silica nanoparticles exhibits the property of iridescence, i.e. the color change depends on the viewing and the illumination angle [see Fig. S1]~\cite{joannopoulos2011photonic, vukusic2003photonic, cong2013preparation, supplemental}.

 \begin{figure*}
 \includegraphics[width=0.8\textwidth]{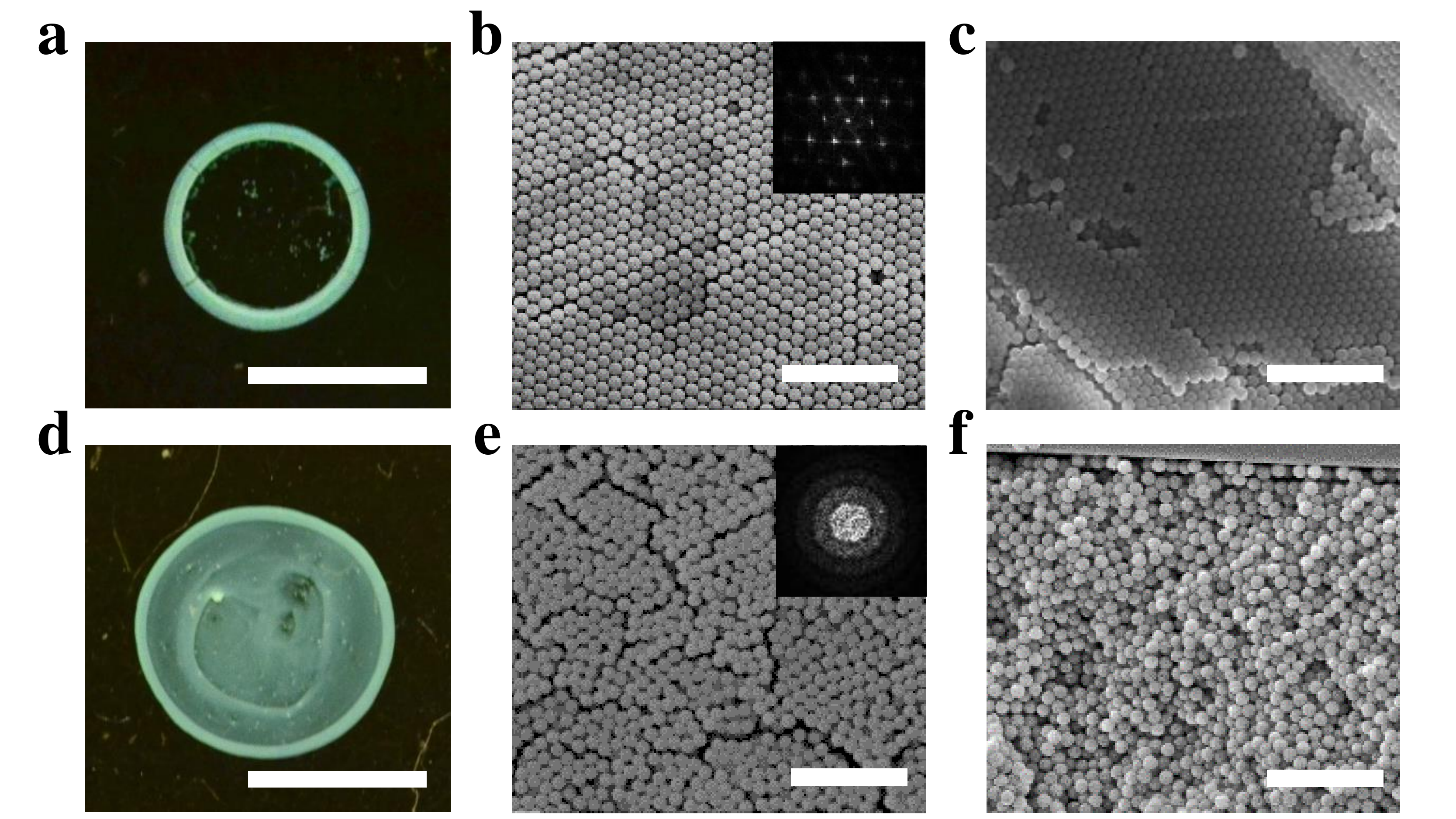}
 \caption{Comparison of dried patterns of different liquid systems containing silica nanoparticles (diameter $d$ = 228 nm) on a cover glass. (a) A dried droplet of DI water with silica nanoparticles at $T_s =$ \SI{25}{\celsius}, where $T_s$ is the temperature of the substrate. (b) and (c), respectively, are surface and cross-sectional SEM images of (a). (d) A dried droplet of a model liquid (DI water + ethanol + PEO) with silica nanoparticles at $T_s =$ \SI{60}{\celsius}. (e) and (f), respectively, are surface and cross-sectional SEM images of (d). The insets in panels (b) and (e) show two-dimensional fast Fourier transforms (2D FFTs) with (a) representative of regular colloidal arrays and (d) representative of short-range-ordered colloidal arrays. The scale bars of (a) and (d) are 2 mm and the scale bars of (b), (c), (e), and (f) are \SI{2}{\micro\meter}.}
 \end{figure*}

In contrast, the final deposit produced from a model liquid (deionized [DI] water + ethanol + polyethylene oxide [PEO]) containing silica (SiO$_2$) nanoparticles ($d$ = 228 nm), which is dried on a substrate with surface temperature, $T_s =$ \SI{60}{\celsius}, exhibits a green and more uniform pattern, as shown in Fig. 1(d). The concentration by volume of PEO and SiO$_2$ in the ethanol-water (50:50 vol. \%) mixture is, respectively, 0.03\% and 1.5\%. Although silica nanoparticles are locally ordered on the surface of the final deposit [Fig. 1(e)], an amorphous array with a short-range order is formed in the entire cross-section except at the surface [Fig. 1(f)]. As the structural color is developed by the internal structure rather than the surface, the deposit is expected to show non-iridescent color~\cite{magkiriadou2012disordered, gotoh2012amorphous, takeoka2009structural,ueno2010soft, lee2016polymeric}. 

In fact, non-iridescent photonic materials that consist of short-range-ordered arrays of monodisperse colloidal nanoparticles have been made by spray coating \cite{takeoka2013production}, centrifugation \cite{kohri2015biomimetic}, or the Leidenfrost effect \cite{lim2014colloidal}. However, these methods focused mainly on the short-range-ordered nanostructures rather than the geometry of the colloidal assemblies at large scales \cite{takeoka2013production, kohri2015biomimetic} and have limitations on patterning and printing technology \cite{takeoka2013production, kohri2015biomimetic, lim2014colloidal}. To resolve these issues, we will focus on both the controlled uniform thickness of the photonic materials and packing structure (ordered versus disordered) of the nanoparticles of the dried coating from model liquids by using simple but robust steps. We expect that our method enables controlled patterning or printing of a non-iridescent photonic material in applications such as ink-jet printing \cite{park2006control, kim2016controlled}, reflection-mode displays \cite{konstantatos2013colloidal}, and other optical devices \cite{choi2009chitosan, liz2006tailoring}.

\section {EXPERIMENTAL PROCEDURE}
\subsection{Silica nanoparticles synthesis}For the experiments, we synthesized monodisperse silica nanoparticles by two steps: silica seed particles are prepared by a two-phase method and then seed particles are grown by the St{\"o}ber method \cite{stober1968controlled, hartlen2008facile}. Silica nanoparticles with diameter $d$ = 195, 228, and 290 nm are prepared to make final deposits with different structural colors.
\subsection{Solution preparation}To design color-selective, non-iridescent materials, the synthesized silica nanoparticles and polyethylene oxide (PEO; $4 \times 10^6$ Da, Sigma-Aldrich, USA) are dissolved in ethanol-water mixtures (50:50 vol. \%). The polymer can be adsorbed onto the silica surface by physisorption. The volume ratio of the ethanol-water mixtures is selected as 1:1 because solvent mixtures at this ratio solvate PEO chains better than single solvents \cite{hammouda2006solvation}. The PEO volume concentration at the surface of the silica nanoparticles is set as the oversaturated concentration at which the remaining PEO in the solution functions as a depletant and makes PEO-coated silica nanoparticles aggregate~\cite{kim2017processable, feng2015re}. With the PEO volume concentration of 1 mg of PEO per 1 m$^2$ surface area of silica particles, we observed silica nanoparticles aggregates in the solution [see supplemental material and movie 5]~\cite{supplemental}, indicating this condition is the oversaturated concentration. It is also important to avoid highly oversaturated PEO concentration. Otherwise, the final deposits lose their structural colors [see Fig. S7]~\cite{supplemental}. Thus, in this study, we decided that the PEO concentration is kept 1 mg of PEO per 1 m$^2$ surface area of silica particles. Continuous mixing of the prepared solution for 24 hours is performed to equilibrate the solutions before use. Density, dynamic viscosity, and surface tension of the initial state of the prepared solutions at $T =$ \SI{25}{\celsius} are 940 kg/m$^3$, 3.0 mPa$\cdot$s, and 29.0 $\pm$ 0.5 mN/m, respectively. At $T =$ \SI{60}{\celsius}, the viscosity of the prepared solution is $2-20$ mPa$\cdot$s showing a shear thinning effect. The diffusion coefficient is estimated using the Stokes-Einstein model, $D = k_BT_{abs}/(6\pi\mu r)$, where $k_B$ is Boltzmann’s constant (1.38 $\times$ $10^{-23}$ J/K), $T_{abs}$ is the absolute temperature (K), $\mu$ is the dynamic viscosity of the solution (Pa$\cdot$s), and $r$ is the radius of the nanoparticles (m). 
The surface tension was measured using a pendant droplet method, which utilized an in-house MATLAB code. The results were validated by comparing with experimental results from a conventional Goniometer (Theta Lite, Biolin Scientific). The viscosities of all solutions were measured using a Rheometer (Anton-Paar MCR 502, Austria) with a CP50-1 geometry. Material weights were measured by a Mettler Toledo XS105 scale.

\subsection{Flow visualization}We performed Particle Tracking Velocimetry (PTV) to measure the flow speed inside the sessile droplet during evaporation. To monitor the flow patterns, \SI{1}{\micro\meter} diameter fluorescent polystyrene particles (Sigma Aldrich) are added to 1 ml prepared solution at a concentration of 0.0025 vol. \%. We observed the particle motion from below with a fluorescent microscope (Nikon Eclipse Ti, Nikon Intensilight C-HGFI). The focal plane is close to the glass surface. The F-doped Tin Oxide (FTO) transparent heater (Solar ceramic, South Korea) is installed between the cover glass and the microscope.
The Stokes number is defined as $St=\tau_p/\tau_l$, where $\tau_p$ is the response time of the particle and $\tau_l$ is the time scale of the fluid motion.; $St$ is the order of magnitude $10^{-10}$ because $\tau_p =(1/18) \rho_p d_p^2/\mu_l \approx$ $10^{-8}$ s and $\tau_l =a/U \approx 100$ s, where $\rho_p$ and $d_p$ are the density (1050 kg/m$^3$) and the diameter of the polystyrene particle, respectively, $\mu_l$ is the mixture viscosity, and $a$ is the droplet radius. $U_M$ is the thermal Marangoni flow speed, $\mathcal{O}$(\SI{10}{\micro\meter}/s), as obtained from the PTV measurement. Therefore, the polystyrene particles can follow the flow inside the droplet perfectly. The terminal velocity of a spherical particle ($u_s=(2/9)(\rho_p-\rho_f )gr_p^2/\rho_l$) can be calculated by balancing the Stokes drag force ($F_D=6\pi\mu_l r_p u$) and the buoyancy force ($F_g=4/3\pi r_p^3 (\rho_p-\rho_f )g$), where $r_p$ is the radius of a spherical particle, $g$ is the gravitational acceleration, and $\rho_f$ is the density of the water-ethanol mixture (940 kg/m$^3$). The terminal velocities of the silica nanoparticles ($r_p=145$ nm and $\rho_p = $2,000 kg/m$^3$) and the polystyrene particles are approximately $1.6 \times10^{-8}$  m/s and $2.0\times10^{-8}$ m/s, respectively, which are negligible compared to the thermal Marangoni flow speed.
\subsection{Camera setup}All of the optical images are taken with a Nikon D5100 under an ambient light. Final deposits of the model liquids with silica nanoparticles on the cover glasses are placed on a black paper to reduce back scattering and enhance structural color. To see the iridescence property, the final deposits are rotated, whereas the camera position is fixed.
\subsection{Packing structure analysis}Images of the particle packing were taken by an XL30 scanning electronic microscope (SEM) at the Andlinger Center, Princeton University. The packing structures of the corresponding SEM images were analyzed by two-dimensional fast Fourier transforms (2D FFT) using MATLAB 2016a.

\section {RESULTS AND DISCUSSION}
\subsection{Experiments for non-iridescent colloidal nanostructures}
A \SI{1}{\micro\liter} droplet of a model liquid (see the experimental procedure and  Sec. S1) containing silica nanoparticles is deposited on a temperature-controlled substrate (a cover glass attached to the hot plate), as sketched in Fig. 2(a)~\cite{supplemental}. The initial contact angle of the sessile droplets of the model liquid on the substrate, with different diameter of silica nanoparticles, are 20 $\pm$ \SI{0.5}{\degree}. During Step 1 of Fig. 2(a), the PEO can adsorb onto the surface of the silica nanoparticles creating a pseudobrush structure \cite{kim2016controlled, cabane1997shear}. At Step 2, during evaporation on a heated substrate, we observed a circulating internal flow in the drop due to thermal Marangoni effects associated with a surface tension gradient \cite{xu2009criterion} after the initial solutal Marangoni flows disappear, which mix the suspended particles, as shown in Fig. 2(b) and supplemental movies 1 and 2 \cite{supplemental}. These effects result in continuous mixing \cite{park2006control, kim2016controlled} of PEO-coated silica nanoparticles and suppression of the coffee-ring pattern \cite{hu2006marangoni, still2012surfactant} during the drying process. We can tune the structural color of the final deposits by selecting the size of the silica nanoparticles, as depicted in Fig. 2(c), and the final deposits have a short-range-ordered packing structure [Fig. 1(f)]. 

We should note that the method introduced in Fig. 2 is robust and simple. Although the experiments were performed in an open lab environment without using a container to precisely control the temperature (24.5 $\pm$ \SI{1.0}{\celsius}) and humidity (18 $\pm$ 3\%) and to isolate the external random effects, this method is highly reproducible under ambient conditions. In addition, compared with other conventional processes of making non-iridescent structures \cite{gotoh2012amorphous, takeoka2013production, garcia2007photonic}, the current approach does not require complicated chemical and physical methods; Firstly, the PEO spontaneously adheres to the silica surface. Secondly, for droplet volumes of \SI{1}{\micro\liter}, the process only takes about 40 seconds to obtain the final deposit using a model liquid. Furthermore, our current drop-casting method can make a nearly uniform thickness over a millimeter-size region, as shown in Fig. 1(d). This is important because the structural colorations by a drop-casting method are usually used to perform color patternings with arrays of small droplets (around a hundred micrometers), which limits use to ink-jet printing \cite{kuang2014inkjet, nam2016inkjet, keller2018inkjet, bai2018large}. However, the large-scale final deposit by the drop-casting method indicates that the process can be possibly applied to not only ink-jet printing, but also a wide variety of printing techniques, such as continuous printing with a dispenser.

 \begin{figure*}
 \includegraphics[width=0.8\textwidth]{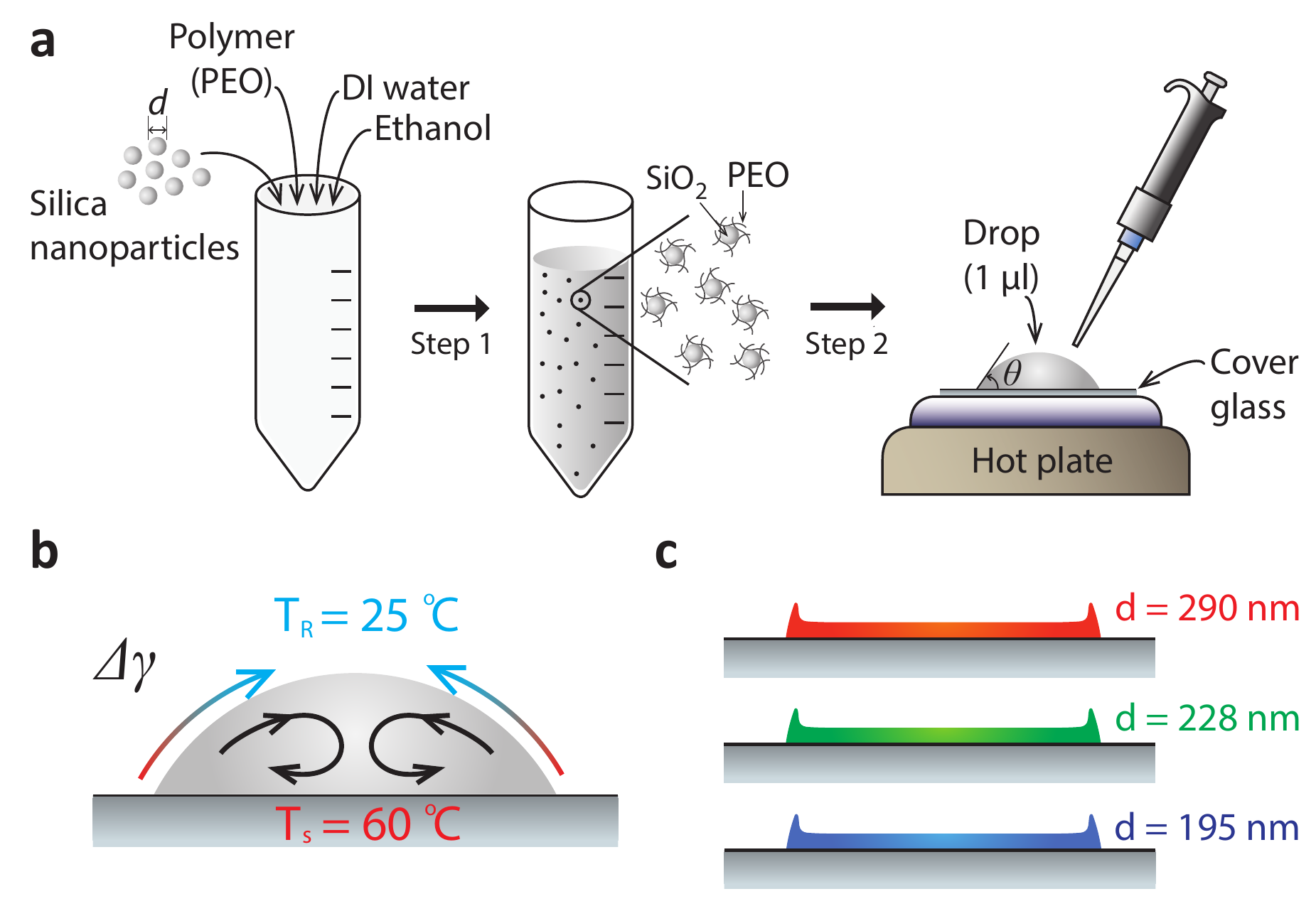}
 \caption{Sketch of experimental procedures. (a) Step 1: Polymer (PEO, molecular weight $4 \times 10^6$ Da) coating silica nanoparticles ($d$ = 195, 228, or 290 nm) in ethanol-water mixtures (50:50 vol. \%). The prepared solution was continuously mixed to equilibrate for 24 hours before use. Step 2: A droplet ($V$ = \SI{1}{\micro\liter}) deposited on a temperature-controlled substrate (a cover glass). The apparent contact angle ($\theta$) is typically \SI{20}{\degree}. (b) Continuous mixing of PEO-coated silica nanoparticles in a droplet driven by thermal Marangoni flow due to a surface tension difference $\Delta\gamma$ along the droplet surface. The black arrows represent a circulating flow pattern driven by thermal Marangoni effects. (c) Outcome: Final deposition pattern and color, which are controlled by the diameter and ordering of the silica nanoparticles.}
 \end{figure*}

\subsection{Thickness uniformity of final deposits}
To examine how the thermal Marangoni effects contribute to the mixing and the uniformity of the final deposition pattern, the dried deposits of ethanol-water mixtures (50:50 vol. \%) containing PEO-coated silica nanoparticles ($d$ = 228 nm) at different substrate temperatures ($T_s$ = (i) \SI{25}{\celsius}, (ii) \SI{40}{\celsius}, and (iii) \SI{60}{\celsius}) are compared, as shown in Fig. 3(a) and 3(b). The thicknesses of the final deposits are measured by profilometry (Leica DCM 3D microscope). For $T_s$ = \SI{25}{\celsius}, most of the silica nanoparticles are deposited at the edge, whereas as $T_s$ increases, silica nanoparticles are deposited more uniformly. Different thickness profiles along the red dashed line of each final deposit of Fig. 3(a) are presented in Fig. 3(b). The case (i) shows a strong ring at the outer edge. However, the deposit thickness becomes almost uniform over a wide area although a ring is still evident at the edge in case (iii). Final deposits with different diameters of silica nanoparticles also show similar deposition patterns and thicknesses, as shown in Fig. S2 \cite{supplemental}.

We monitor the fluid motion by seeding \SI{1}{\micro\meter} diameter fluorescent polystyrene particles in the model liquid. When the model liquid droplet is deposited and dries on a heated substrate, initially, solutal Marangoni flows dominate over the thermal Marangoni flow. In this system, the solutal Marangoni mixing flow continues for about 15 seconds until most of the ethanol has evaporated, which is consistent with previous studies \cite{kim2016controlled}, and subsequently a thermal Marangoni flow was observed (supplemental movies 1 and 2) \cite{supplemental}. As a result, two sequential Marangoni effects occur and suppress formation of a coffee ring. Then, while the thermal Marangoni effect occurs, continuous de-pinning of the contact line is observed. 

 \begin{SCfigure*}
 \centering
 \caption{Thermal Marangoni effects on the final deposition pattern. (a) Comparison of the dried mark of an ethanol-water mixture (50:50 vol. \%) containing PEO-coated silica nanoparticles ($d$ = 290 nm) at different substrate temperatures, (i) $T_s$ = \SI{25}{\celsius}, (ii) $T_s$ = \SI{40}{\celsius}, and (iii) $T_s$ = \SI{60}{\celsius}. The scale bars are 2 mm. (b) Comparison of thickness profiles along the red dashed line of (a); the light gray solid line, $T_s$ = \SI{25}{\celsius}, the dark gray solid line, $T_s$ = \SI{40}{\celsius}, and the black solid line, $T_s$ = \SI{60}{\celsius}. The thickness was measured by profilometry. (c) Temporal plot of temperature at the periphery and center of the droplet at $T_s$ = \SI{60}{\celsius} during evaporation. The temperature was measured using an IR camera. (d) Time evolution of the thermal Marangoni flow speed ($U_M$) and droplet surface shrinkage speed ($U_h$) during the droplet evaporation at $T_s$ = \SI{60}{\celsius}. The droplet apex shrinkage speed $U_h$ is obtained from a side view measurement and has a standard deviation error, 7\%, which is obtained from multiple experiments.}
 \includegraphics[width=0.72\textwidth]{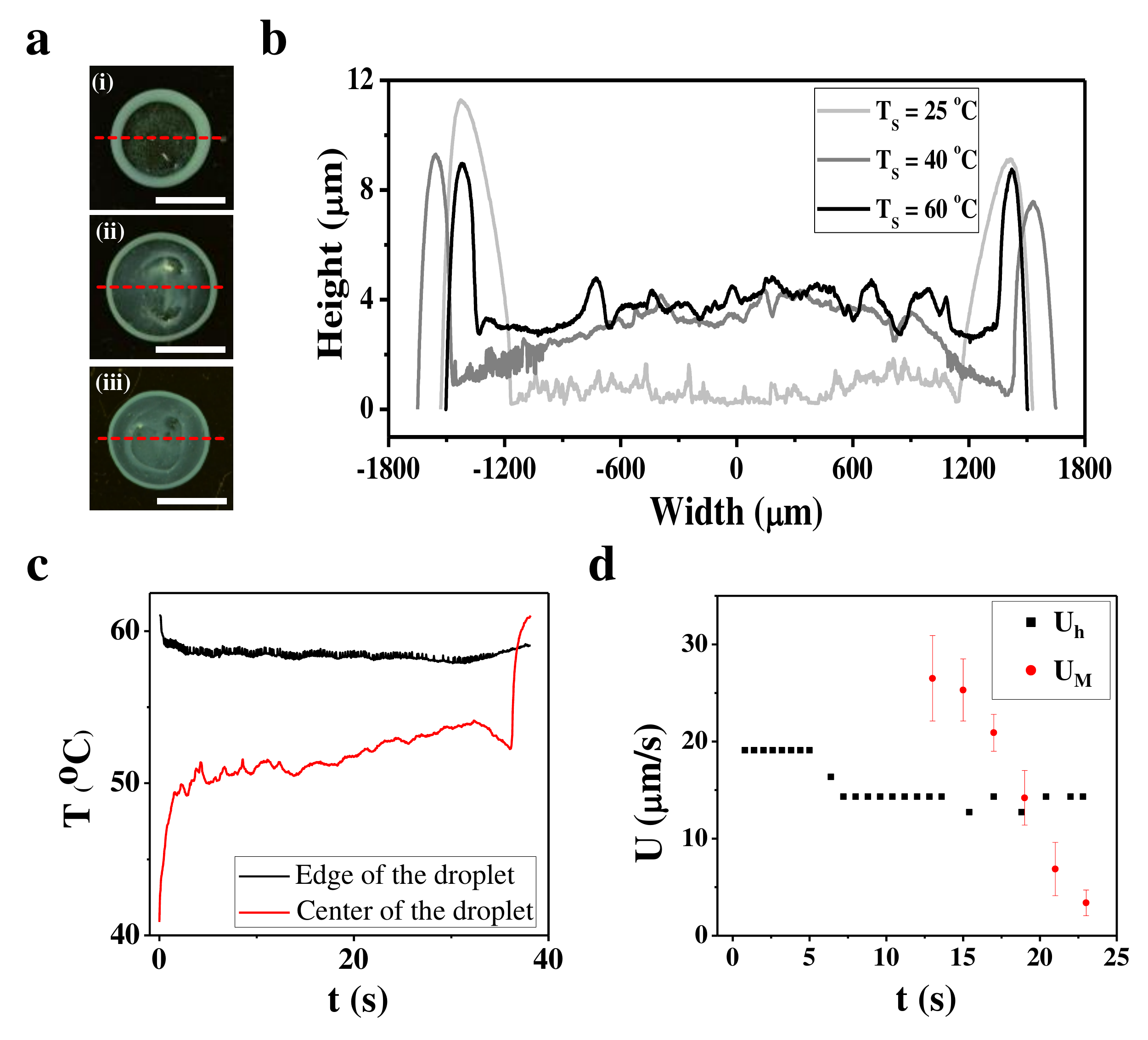} 
 \end{SCfigure*}

To explain effects of the temperature on the final deposition pattern, we measured the temperature distribution of the surface of the model liquid droplet using an IR camera (FLIR A300). We obtained the temperature gradient from the edge to the center of the droplet throughout the evaporation process (see Fig. 3(c) and Fig. S4). This temperature gradient along the droplet surface leads to a surface tension gradient along the droplet surface and the corresponding thermal Marangoni flow continuously mixes the PEO-coated silica nanoparticles. To examine the dominant effect between the temperature-dependent surface-tension-driven flows and viscous effects due to flow, we consider the Marangoni number, $Ma\equiv-(d\gamma/dT) (H\Delta T/\mu \kappa)$ where $d\gamma/dT$ is the surface tension change with temperature, $\mu$ is the dynamic viscosity, $\kappa$ is the thermal diffusivity of the solution, and $H$ is the initial height of a droplet. Then, the Marangoni number is calculated for a water droplet assuming ethanol is almost completely evaporated: at $T_s$ = \SI{60}{\celsius}, $Ma = 3.4\times10^4$, where $d\gamma/dT$ = -0.17 mN/m$\cdot^\circ$C \cite{vargaftik1983international}, $\mu = 4.7 \times10^{-4}$ Pa$\cdot$s \cite{sengers1986improved}, $\kappa=1.6 \times10^{-7}$ m$^2/$s \cite{sengers1986improved, marsh1987recommended}, $\Delta T\approx 10 ^\circ$C, and $H=0.26$ mm; the concentration of the nanoparticles does not significantly change the physical properties of water. The magnitude of the Marangoni number ($Ma \gg$ 1) shows that the thermal Marangoni effects, due to surface tension gradients produced by a temperature difference, are much stronger than viscous effects, and so drive the circulating flow as sketched in Fig. 2(b) \cite{girard2008effect}.

\begin{figure*}
 \includegraphics[width=0.8\textwidth]{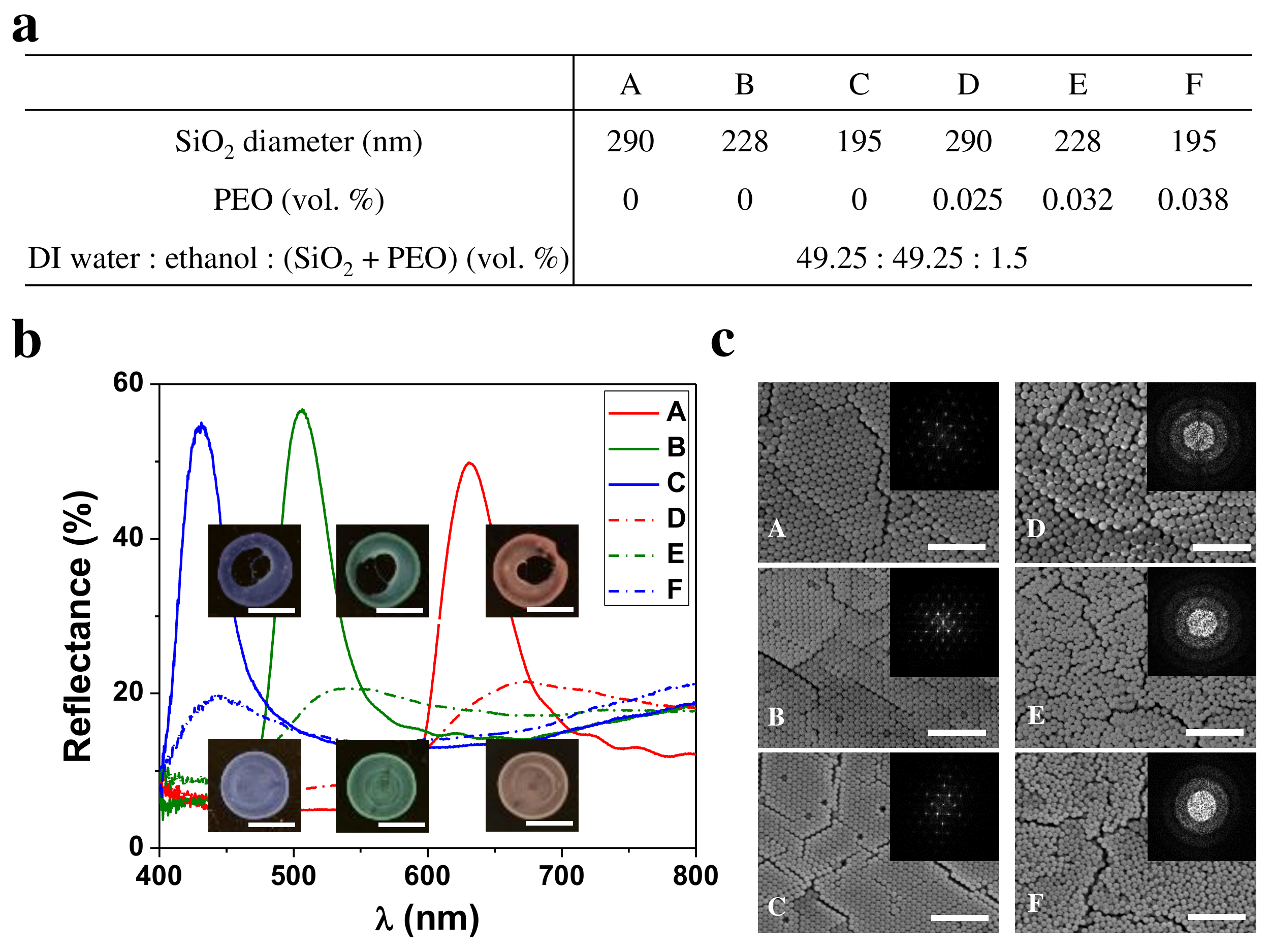}
 \caption{Polymer effects on the packing structure of silica nanoparticles. (a) Table of prepared solution compositions. For the solutions D, E and F, 1 mg/m$^2$ of PEO polymer to be adsorbed on the surface of the silica nanoparticles are added. (b) Reflectance spectra of dried deposits set by the solutions listed in (a). Insets are the final deposition patterns of different cases. Scale bars in the insets are 2 mm. (c) SEM images from the final deposits of (A-F) are taken after droplets are dried on a \SI{60}{\celsius} substrate. Scale bars are \SI{2}{\micro\meter}. The insets of each of the SEM images in panel (c) present 2D FFTs. \label{bbb}}
 \end{figure*}

Furthermore, the heated substrate enhances the evaporation rate of the deposited droplet, which increases the speed of shrinkage of the droplet surface ($U_h$). If the shrinkage speed is faster than the particle migration speed in the radial direction, the particles can be possibly captured at the droplet surface or trapped on the substrate by collapsing the droplet surface, which can induce a relatively uniform coating \cite{kang2016alternative, li2016rate}. We measure two speeds for the thermal Marangoni flow along the droplet radius ($U_M$) and the interface shrinking rate of the apex of the droplet ($U_h$), as shown in Fig. 3(d) and supplemental movies 2 and 4) \cite{supplemental}. We observe that the speeds of the two processes cross over at a late time period ($t \simeq 19 $ s). While $U_M > U_h$, the thermal Marangoni stresses contribute to suppress the coffee-ring stain. Later, if $U_M < U_h$, the fast interface shrinkage speed causes capture of more particles at the center of the droplet.

\subsection{Packing nanostructure of final deposits}
To observe the effects of polymer on the final deposits of the silica nanoparticles, one set of three different suspensions without PEO (A-C) and the other set with PEO (D-F) are prepared [see Fig. 4(a)]. For each case, we deposited a droplet ($V$ $=$ \SI{1}{\micro\liter}) on the \SI{60}{\celsius} cover glass. Reflectance spectra of the final deposits of suspensions (A-F) are measured by placing a fiber-type spectrometer (Ocean Optics LS-1-LL) normal to the surface of the deposits on the cover glass. From the spectrometry measurements, the maximum wavelengths ($\lambda_{peak}$) of the spectra of the final deposits without PEO (A-C) are, respectively, 632, 506, and 431 nm, whereas those with PEO are, respectively, 670, 534, and 446 nm [Fig. 4(b)]. The results with PEO have weaker, broader, and red-shifted reflectance peaks, which imply that the structures with PEO are more disordered than the structures without PEO.

SEM images taken at the edge of the final deposits are presented in Fig. 4(c): the pure silica nanoparticles are almost close-packed in hexagonal arrays at the surface of the final patterns from solutions A, B, and C, which contain no PEO, whereas silica nanoparticles are packed in more disordered arrays for solutions D, E, and F, which contain PEO. Cross-sectional images of the final patterns taken at the edge (Fig. S5) and surface images of the final pattern taken at edge and center (Fig. S6) also support the explanation for Fig. 4(c)~\cite{supplemental}. Additionally, the volume fractions of the packed structures with PEO are examined to learn whether the packing is disordered or not. We assumed that the final deposits are axisymmetric and the volume fraction of nanoparticles is estimated by integrating the thickness profiles [see Sec. S4]~\cite{supplemental}. In the case of an ethanol-water mixture (50:50 vol.\%) containing PEO-coated silica nanoparticles ($d$ = 228 nm) at different $T_s$, e.g. (i) $T_s$ = \SI{25}{\celsius}, (ii) \SI{40}{\celsius}, and (iii) \SI{60}{\celsius}, the volume fractions of silica nanoparticles are approximately 0.56-0.62, which are consistent with a typical volume fraction of disordered close-packing; the densest random packing case has a volume fraction about 0.64 \cite{scott1969density}.

 \begin{figure*}
 \includegraphics[width=0.9\textwidth]{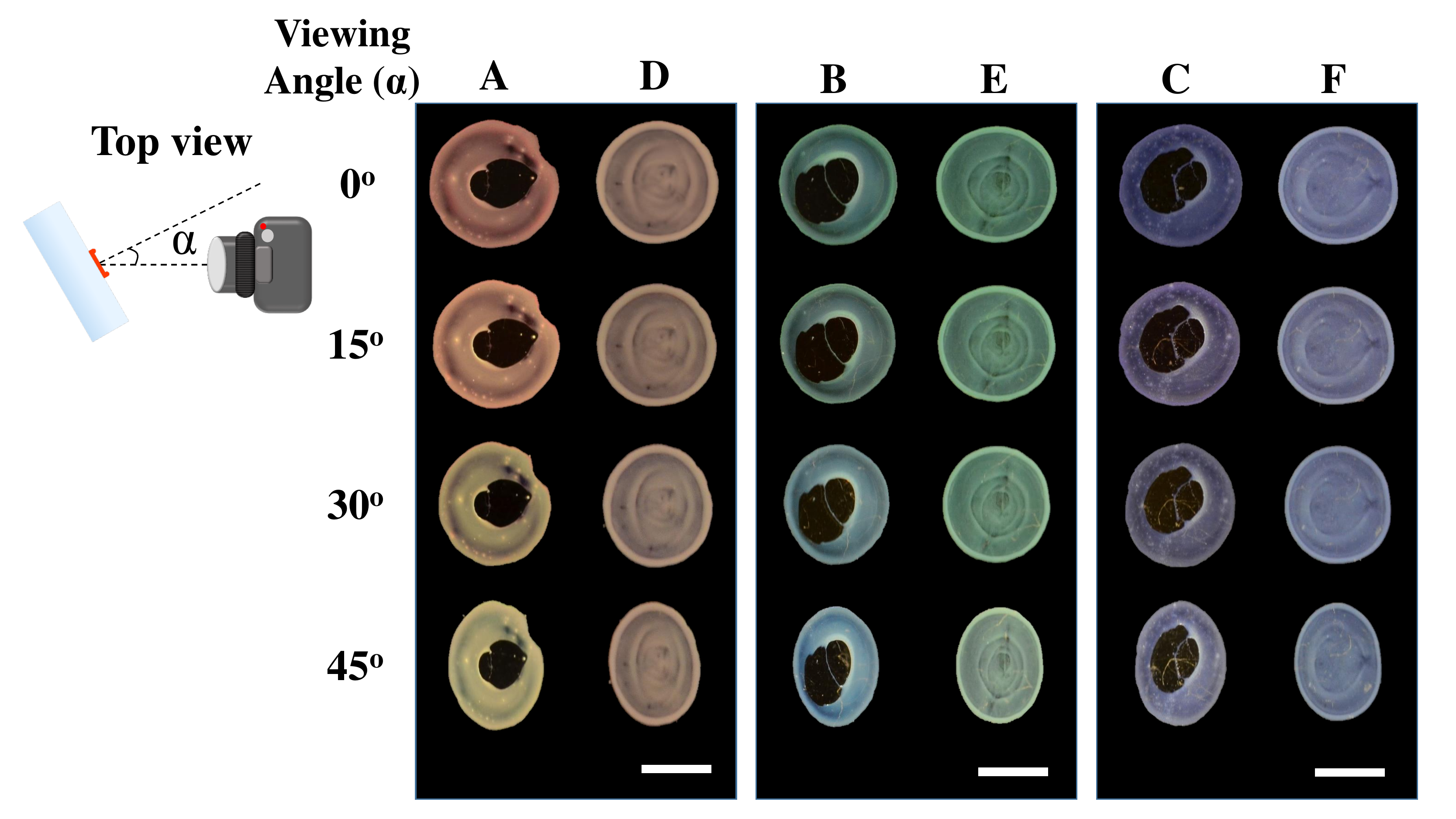}
 \caption{Non-iridescent properties of dried patterns of the model liquids. Photographs of deposited patterns with three different sizes of silica nanoparticles, which are originally composed of (left column) silica + water + ethanol (solution A, B, and C) and (right column) silica + water + ethanol + PEO (solution D, E, and F) where the viewing angle ($\alpha$) is varied. The pictures are taken under ambient light. Scale bars are 2 mm. \label{bbb}}
 \end{figure*}

At the initial stage of the evaporation of the solutions D, E, and F in Fig. 4(a), PEO-coated silica nanoparticles are well-dispersed in the solution \cite{cabane1997shear}. The concentration of any remaining PEO in the bulk phase, which does not adsorb on the silica surface, functions as a depletant~\cite{kim2017processable, feng2015re}. This depletion attraction makes silica nanoparticles aggregate. As the silica nanoparticles physically contact each other, friction forces may occur between the PEO brushes adsorbed on the silica nanoparticles or on the glass substrate. Due to the depletion attraction and friction forces between polymer brushes, silica nanoparticles remain stuck in the position once they are in contact. These interactions cause the final deposits to be short-range-ordered. To examine effects of the initial concentration of PEO, we performed additional experiments that showed that an increase of the initial concentration of PEO makes a stronger depletion interaction and the resulting aggregates will be further disordered (supplemental movie 5 and Fig. S7) \cite{supplemental}.

We can also estimate the packing results by comparing two time scales: the time scale for particle motion by diffusion and the time scale for the hydrodynamic effect associated with the surface shrinkage speed of the droplet. The diffusive time scale can be estimated as $\tau_D = L^2/D$ where $L$ is the typical distance between particles in the evaporating droplet and $D \approx 10 ^{- 12}$ m$^2$/s is the diffusion coefficient defined from the Stokes-Einstein model (see Experimental procedure). Also, $\tau_h = L/U_h$ is the hydrodynamic time scale for the surface shrinkage speed where $U_h = 10^{-5}$ m/s. In this case, $L$ is always larger than the particle diameter $d$, so that it is always true that $\tau_D/ \tau_h > 1$. Therefore, the rapid shrinkage speed of the droplet surface contributes to the creation of a short-range-ordered structure.

\subsection{Non-iridescent properties of final deposits}
Finally, we show the non-iridescent structural colors from the short-range-ordered close-packed nanoparticles for $d$ = 195, 228, and 290 nm. To test the non-iridescent properties, we observed the final deposits of samples from the solutions A-F (Fig. 4) at different viewing angles ($\alpha$) under an ambient light (see the schematic in Fig. 5). In the absence of PEO, the structural colors of the final deposits of silica + DI water + ethanol are blue shifted as $\alpha$ is changed from \SI{0}{\degree} to \SI{45}{\degree}. However, Fig. 5 and Fig. S8 show that the final deposits of liquids with silica nanoparticles and PEO remain unchanged with changes in the viewing angle, due to the isotropic short-range-ordered packing structure of the nanoparticles. As a result, we obtain a nearly uniform non-iridescent nanostructure with our model liquid.

\section {CONCLUSION}
In this study, we have created non-iridescent self-assembled nanostructures with a nearly uniform thickness by simply evaporating suspension droplets of a model liquid on a heated substrate; the model liquid consists of PEO and PEO-coated silica nanoparticles suspended in a water-ethanol mixture. The thickness uniformity of the final deposits after drying can be controlled by the temperature of the substrate, which induces thermal Marangoni stresses along the surface of the droplet. This Marangoni flow prevents the colloidal nanoparticles from forming a coffee ring and makes relatively uniform thickness profiles along the final deposits. Furthermore, we have changed the optical property of the final deposits by adding the polymer in the droplet solution. PEO physically adsorbs on the surface of the silica nanoparticles and the remaining PEO in the solution acts as a depletant. These effects prevent silica nanoparticles from forming a face-centered-cubic structure. This short-range-ordered array of nanoparticles is different from a photonic crystal and exhibits a non-iridescent property. We believe that our simple and highly reproducible drop-casting method can be applied to a wide range of printing technologies, such as ink-jet printing or direct writing. In addition, this method can be used in light filtering systems, e.g. lenses and glass windows.

\section* {ACKNOWLEDGEMENTS}
We thank J. Nunes at Princeton University for useful conversations and for the initial preliminary SEM measurement. We thank B.J. Lee at KAIST for sharing the temperature measurement device. We also thank N. Yao from Andlinger Center in Princeton University and S. Kelly from Carl Zeiss for 3D X-ray microscope images by a Zeiss Xradia 810 Ultra. This work was supported by the Global Research Laboratory (GRL) Program and Young Researcher Program through the National Research Foundation (NRF) of Korea funded by the Ministry of Science and ICT (NRF-2015K1A1A2033054 and NRF-2018R1C1B6004190, respectively) and the Settlement Research Funds for newly hired tenure track faculty from KAIST (G04170022).


\end{document}